\documentclass[12pt]{article}
\usepackage{amssymb,amsmath}
\usepackage[square,
comma,
sort&compress]{natbib}
\include{epsf}

\usepackage{graphicx}
\usepackage{amsfonts}

\newcommand{\e}{{\rm e}}
\newcommand{\ii}{{\rm i}}
\newcommand{\dd}{{\rm d}}
\newcommand{\eqn}[1]{(\ref{#1})}

\def\appendix#1{\addtocounter{section}{1}\setcounter{equation}{0}
\renewcommand{\thesection}{\Alph{section}}
\section*{
\thesection\protect\indent \parbox[t]{11.715cm} {#1}}
\addcontentsline{toc}{section}{Appendix\thesection\ \ \ #1} }
\newcommand{\real}{{\mathbb R}} 
\def\bra#1{\left\langle #1\right|}
\def\ket#1{\left| #1\right\rangle}
\def\hs#1#2{\left\langle #1\right.\left| #2\right\rangle}
\def\hstar#1#2{\left\langle #1\stackrel{\star}{|} #2\right\rangle}

\newcommand{\be}{\begin{equation}}
\newcommand{\ee}{\end{equation}}
\newcommand{\beq}{\begin{equation}}
\newcommand{\eeq}{\end{equation}}
\newcommand{\bea}{\begin{eqnarray}}
\newcommand{\eea}{\end{eqnarray}}
\def\beqa{\begin{eqnarray}}
\def\eeqa{\end{eqnarray}}
\def\nn{\nonumber}
\newcommand{\del}{\partial}

\newcommand{\eq}{\begin{equation}}
\newcommand{\eqa}{\begin{eqnarray}}
\newcommand{\en}{\end{equation}}
\newcommand{\ena}{\end{eqnarray}}

\newcommand{\starmoy}{\star_M}
\newcommand{\starvor}{\star_V}
\def\hsstar#1#2{\left\langle #1\stackrel{\star}{\big|} #2\right\rangle}
\def\hsstarmoy#1#2{\left\langle #1\stackrel{\starmoy}{\big|} #2\right\rangle}
\def\hsstarvor#1#2{\left\langle #1\stackrel{\starvor}{\big|} #2\right\rangle}
\newcommand{\fase}[2]{#1 \bullet #2}

\begin{document}
\begin{titlepage}
\begin{flushright}

\baselineskip=12pt
DSF--19--2008\\
\hfill{ }\\
September 2008
\end{flushright}

\begin{center}

\baselineskip=24pt

{\Large\bf Twisted Noncommutative Field Theory with the Wick-Voros
and  Moyal Products}

\baselineskip=14pt

\vspace{1cm}

{\bf Salvatore Galluccio, Fedele Lizzi and Patrizia Vitale}
\\[6mm]
{\it Dipartimento di Scienze Fisiche, Universit\`{a}
di Napoli {\sl Federico II}}\\ and {\it INFN, Sezione di Napoli}\\
{\it Monte S.~Angelo, Via Cintia, 80126 Napoli, Italy}\\{\small\tt
salvatore.galluccio@na.infn.it, fedele.lizzi@na.infn.it,
patrizia.vitale@na.infn.it}
\\[10mm]

\end{center}

\vskip 2 cm

\begin{abstract}
We present a comparison of the noncommutative field theories built
using two different star products: Moyal and Wick-Voros (or
normally ordered). For the latter we discuss both the classical
and the quantum field theory in the quartic potential case, and
calculate the Green's functions up to one loop, for the two and
four points cases. We compare the two theories in the context of
the noncommutative geometry determined by a Drinfeld twist, and
the comparison is made at the level of Green's functions and
S-matrix. We find that while the Green's functions are different
for the two theories, the S-matrix is the same in both cases, and
is different from the commutative case.
\end{abstract}

\end{titlepage}

\section{Introduction}
It is likely that at short distances spacetime has to be described
by different geometrical structures, and that the very concept of
point and localizability may no longer be adequate. This is one of
the main motivations for the introduction of noncommutative
geometry~\cite{connes, landi, ticos}. The simplest kind of
noncommutative geometry is the so called \emph{canonical}
one~\cite{DoplicherFredenhagenRoberts,Doplicher}. What is usually
done for the construction of a field theory on a noncommutative
space is to deform the product among functions (and hence among
fields) with the introduction of a noncommutative $\star$ product,
so that for the coordinate functions one has
\be
[x^i,x^j]_\star\equiv x^i\star x^j-x^j\star x^i=\ii\theta^{ij}.
\label{commutatorx}
\ee
In the simplest case  $\theta^{ij}$ is constant, i.e.\ it does not
depend on the $x$'s. The choice of the $\star$ product compatible
with \eqn{commutatorx} is not unique, in the following we will
introduce two different products, the Moyal~\cite{Gronewold,Moyal}
and Wick-Voros~\cite{Bayen, Voros, BordemannWaldmann1,
BordemannWaldmann2} and compare their ``physical predictions".

There are several reasons to consider field theories on a
noncommutative space equipped with the standard canonical
noncommutativity, ranging from intrinsic motivations to the
localizability of
events~\cite{DoplicherFredenhagenRoberts,Doplicher} to string theory
~\cite{SeibergWitten} to constructive field
theories~\cite{Rivasseau}. Field theories on noncommutative spaces
have interesting renormalization properties~\cite{GrosseSteinacker,
GrosseWulkenhaar}. For a review see~\cite{Szabo, DouglasNekrasov},
their references and their citations. What we will compare are field
theories in which the product among fields is substituted by the two
different $\star$ products. This leads to an action in which
arbitrary degree derivatives of the fields are present, as a series
in $\theta$. Written in terms of derivatives the two actions with
the Moyal and Wick-Voros products are different. There is however a
map which renders equivalent the algebras generated by the two
products. Field theory with the Wick-Voros product has been
discussed in~\cite{ChaichianDemichevPresnajder} as regularizing
model, their conclusion (that ultraviolet divergences persist) is in
agreement with ours.

This paper originates from the consideration that one can reason
in two ways: one point of view is to say that what counts is the
noncommutative structure of spacetime, and the $\star$ product is
just a way to express this intimate structure, and therefore one
chooses the most convenient product. As long as one is describing
the same field theory, the results should be the same, a fact
noted already in~\cite{HLS-J}. Another view is to claim total lack
of interest in the noncommutativity of spacetime. What counts is
the fact that one has a field theory on ordinary spacetime, whose
action contains an infinity of derivatives of arbitrary order.
With this second point of view one would not in principle expect
the same physical results from the two theories. In this paper we
calculate the Green's functions of the Wick-Voros field theory,
and found them to be different from the Moyal case. This leads to
a contradiction. We will see that the contradiction is only
apparent. Green's functions are not observable quantities, what is
observable is the S-matrix.

Discussions of the properties of the S-matrix often go together
with the issue of Poincar\'e invariance. Relation
\eqn{commutatorx} is not Poincar\'e invariant, and this casts
doubts on its being fundamental. It is however possible to build a
theory which is invariant under a \emph{deformation} of the
Poincar\'e Lie algebra, so that the theory becomes a
\emph{twisted} theory. This theory has a symmetry described by a
noncommutative, non cocommutative Hopf algebra. In particular the
kind of noncommutativity described by the two $\star$ products is
the one generated by a \emph{twist}~\cite{Drinfeld1, Drinfeld2,
Oeckl}. Then the theory has  a twisted Poincar\'e
symmetry~\cite{Helsinki, Wess, aletWess}.

The presence of a twist forces to reconsider all of the steps in a
field theory, which has to be built in a coherent ``twisted'' way.
We will see that there is equivalence between the two theories at
the very end, where by ``very end'' we mean the calculation of the
S-matrix. Prior to this, vertex, propagators and Green's functions
are in fact different. Moreover the equivalence is only obtained if
a consistent procedure of \emph{twisting all products} is applied.
 In this way the Poincar\'e
symmetry, which appears to be broken in~\eqn{commutatorx}, is
preserved, albeit in a deformed way, as a noncommutative, non
cocommutative Hopf algebra.

There is some ambiguity in the issue of twisting, and some results
have been somewhat
 controversial~\cite{Vassilevich, Balspinstat, Balsmatrix, Tureanu, twistandshout, Koreans, Zhan, WessFiore, Fiore, ChaichianSalminenTureanu}. In an optimal world one should let
experiments resolve these ambiguities. Unfortunately the
noncommutative structure of spacetime is not yet mature for a
confrontation with experiments at such a level. What we do in this paper is to use the
field theories built with the Wick-Voros and Moyal products to
check each other. We will see that using a consistent twisting
procedure we obtain that at the level of S-matrix the two
theories are equivalent. This gives us the indication on the
procedure to follow  for noncommutative  theories coming from a twist.

In this paper we will consider exclusively spatial
noncommutativity, i.\e.\ time is a commuting variable. The matrix
$\theta$ therefore is of the form
\be
\theta^{ij}=\theta \varepsilon^{ij}
\ee
with $\varepsilon$ the antisymmetric tensor of order two.

The paper is organized as follows. In Section~2 we introduce the
two products. In section~\ref{se:classical} we discuss the
classical free field theory for the Wick-Voros product. In
section~\ref{se:compare} we calculate the Green's functions for
the two theories for the two and four point case to one loop, and
compare the two cases. In Sect.~\ref{se:twistncg} we describe the
two products as twisted noncommutative geometries. In
section~\ref{se:twistproducts} we describe the relevant twisted
products which we then use in Sect.~\ref{se:smatrix} to calculate
the S-matrix. A final short section contains the conclusions.

\section{The Wick-Voros and Moyal Products}
\setcounter{equation}{0}

In this section we describe in a comparative way the two $\star$
products we are using in this paper. The most well known product
is the  Moyal product~\cite{Gronewold,Moyal}
\be f(\vec
x)\starmoy g(\vec x)=f(\vec x)
\e^{\frac{\ii}{2}\theta^{ij}\overleftarrow{\del_i}\overrightarrow{\del_j}}
g(\vec x)
\ee
where the operator $\overleftarrow{\del_i}$ (resp.\
$\overrightarrow{\del_j}$) acts on the left (resp.\ the right). This
product comes from a Weyl map which associates to a function on the
plane an operator according to:
\be
\hat\Omega_{\rm M}(f)=\frac1{2\pi}\int\dd^2\eta \tilde
f(\eta_1,\eta_2) \e^{\ii \theta_{ij}\hat X^i \eta^j} \label{Weylmap}
\ee
where $\tilde f$ is the symplectic Fourier transform of the function
$f$:
\be
\tilde f =\frac1{\theta\pi}\int\dd^2x \tilde f(x^1,x^2) \e^{-\ii
\theta_{ij}x^i \eta^j}
\ee
$\theta_{ij}$ is the inverse of $\theta^{ij}$, and the $\hat X$ are
operators which satisfy the commutation relation
\be
[\hat X^i,\hat X^j]=\ii\theta^{ij} \label{commX}
\ee
It is useful to think of the operators $\hat X$ in two dimensions in
an abstract way, without reference to spacetime and define them
 as
\bea
\hat X_1 = \frac{\hat a + \hat a^\dagger}{\sqrt 2}\nonumber\\
\hat X_2 = \frac{\hat a - \hat a^\dagger}{\ii\sqrt 2} \label{defX}
\eea
and $\hat a$ is an operator which we define on a certain basis by
its components as
\be
\hat a_{nm}=\sqrt{\theta n}\delta_{m,n+1} \label{defa}
\ee
with $m,n\geq 0$. Of course we are using the analogy of the
commutation relations~\eqn{commX} with the usual quantum mechanical
commutation relations, and using the $\ket{n}$'s as a convenient
basis.

The Moyal product is then defined as
\be
f\starmoy g = \Omega^{-1}_{\rm M}\left(\hat\Omega_{\rm M}(f)\hat
\Omega_{\rm M}(g)\right) \label{moyal}
\ee
From this expression is not difficult (see for
example~\cite{Zampinithesis}) to obtain integral expressions for
the product, a few of which are collected
in~\cite[Appendix]{Selene}. The standard expression is then an
asymptotic expansion of the integral
expressions~\cite{VarillyGraciaBondia}.

One important property of the Moyal product is that
\be
\int\dd^2 x f\starmoy g=\int\dd^2 x f  g \label{Moyintegral}
\ee
and obviously
\be
x^1\starmoy x^2- x^2 \starmoy x^1=\ii \theta
\ee

We now proceed to the definition of the Wick-Voros product. For the
following it is useful to consider the space as a complex plane
defining:
\be
z_\pm=\frac{x^1\pm\ii x^2}{\sqrt{2}}  \label{defz}
\ee
where of course $z_+^*=z_-$. With this substitution we define the
Wick-Voros product as
\be
f\starvor g=\sum_n \left(\frac{\theta^n}{n!}\right) \del_+^n f
\del_-^n g = f
\e^{\theta\overleftarrow{\del_+}\overrightarrow{\del-}}g\label{vorosprod}
\ee
where
\be
\del_\pm=\frac{\del}{\del
z_\pm}=\frac{1}{\sqrt2}\left(\frac\del{\del
{x^1}}\mp\ii\frac\del{\del {x^2}}\right)
\ee
Notice that the Moyal product \eqn{moyal} may be rewritten in these
coordinates as
\be f\starmoy g=f
\e^{\frac{\theta}{2}(\overleftarrow{\del_+}\overrightarrow{\del_-}-
\overleftarrow{\del_-}\overrightarrow{\del_+})} g
\ee
It results
\bea
z_+\starvor  z_-&=& z_+z_- +\theta\nonumber\\
z_-\starvor  z_+ &=& z_+z_-
\eea
and therefore
\be
[z_+,z_-]_{\starvor}=\theta
\ee
Going back to the $x$'s, it is possible to see that this relation
gives rise again to the standard commutator among the $x$'s:
\be
x^1\starvor x^2- x^2 \starvor x^1=\ii \theta
\ee

With the $z_\pm$  coordinates the Laplacian and the d'Alembertian
are respectively $\nabla^2=2 \del_+\del_-$ and
$\Box=\del_0^2-\nabla^2$. The integral on the plane is still a
trace, but the strong condition of~\eqn{Moyintegral} is not valid
anymore:
\be
\int \dd^2 z  f\starvor g=\int \dd^2 z  g\starvor f\neq \int \dd^2
z f  g
\ee
where by $\dd^2z$ we mean the usual measure on the plane $\dd z_+\dd
z_-$. We will also use the notation
\be
k_\pm=\frac{k_1\pm\ii k_2}{\sqrt{2}}
\ee
for a generic vector $\vec k$.

The Wick-Voros and Moyal products can be cast in the same general
framework in that they are both coming from a generalised  ``Weyl
map". More precisely, as we saw in~\eqn{Weylmap} the Moyal product
comes from a map which associates operators to functions, with
symmetric ordering. The Wick-Voros product comes from a similar map,
a \emph{weighted} Weyl map as follows:
\be
\hat\Omega_V (f)= \frac1{2\pi}\int\dd^2 \eta \tilde f(\eta,\bar \eta) \e^{
\theta \eta a^\dag} \e^{ -\theta \bar \eta a} \label{weiWeylmap}
\ee
An equivalent way to associate the operators
$\hat\Omega_V (f)$ to a function $f=\sum_{mn}f_{mn} z_+^m z_-^n$
analytic on the plane is:
\be
\hat\Omega_V(f)=\sum_{mn}f_{mn} \hat a^{\dagger m} \hat a^n
\label{weightedmap}
\ee
where $\hat a $ has been defined in~\eqn{defa}. Thus effectively
the map~\eqn{weightedmap} corresponds to the normal (or Wick)
ordering (and is sometimes called normal ordered product). In this
sense the two maps correspond to two different quantization
procedures~ (see for example~\cite{Cohen, Castellani}).

\section{Classical Free Field Theory \label{se:classical}}
\setcounter{equation}{0}

Although the main interest of this paper is in the interacting
quantum field theory, we start the discussion from the classical
free case. In this section we discuss the kind of field theory one
obtains from a deformation of the free Klein-Gordon action based
on the Wick-Voros product. Such analysis is unnecessary in the
Moyal case, because in that case the action, being quadratic in
the fields,  is the same as in the commutative case.

We consider a field theory described by an action which is a
Wick-Voros deformation of a scalar field theory action, obtained
inserting the star product. Consider a classical free theory and
its action, Lagrangian density and Lagrangian defined as:
\be
S_0=\int \dd t L_0=\int \dd t \dd^2z\,{\cal L}_0=\int \dd t
\dd^2z\, \frac12\left(\del_\mu\varphi\starvor\del_\mu\varphi
-m^2\varphi\starvor\varphi\right) \label{freeaction}
\ee
With the help of \eqn{vorosprod} it may be rewritten as
\bea
S_0&=&\int \dd t
\dd^2z\,\frac12\left(\del_\mu\varphi\e^{\theta\overleftarrow{\del_+}\overrightarrow{\del-}}\del_\mu\varphi
-m^2\varphi\e^{\theta\overleftarrow{\del_+}\overrightarrow{\del-}}\varphi\right)\nonumber\\
&=&\int \dd t \dd^2z\, \frac12
\varphi\left[\e^{-\frac\theta2\nabla^2}(-\del_\mu^2-m^2)\right]\varphi
\label{freac}
\eea
This is a theory which contains an infinite number of the
derivatives of the fields, and in principle even the Cauchy
problem would not be well defined. This appears to be the biggest
difference with respect to the noncommutative field theory defined
via the Moyal product. In the latter case the action being the
same as in the commutative case,  the solution of the free theory
is still given by plane waves, and upon quantization the
propagator is the same as in the commutative case. This time
instead the action is different, the theory is non local as it
contains derivatives of arbitrary order.

The two products are equivalent in a precise technical sense:
there is an invertible map~\cite{Zachos, AlexanianPinzulStern}
\be
T(f)=\sum_n \theta^n t_n(f) \label{equivalencemap}
\ee
with the $t_n$ differential operators, such that
\be
T(f\starmoy g)=T(f)\starvor T(g)
\ee
where
\be
T=\e^{\frac\theta4\nabla^2}\label{Tmoyvor}
\ee
Therefore the two products define the same deformed algebra. This
is certainly true if we consider functions as  formal power series
in the generators. The issue can be more complicated in the realm
of $C^*$-algebras. Starting from the same set of functions the
completion in the supremum norm of the two products could in
principle be different.

The fact that the algebra is the same does not mean that the two
deformations of an action are the same. Therefore let us map the
free action $S_0$~\eqn{freeaction} written with the Wick-Voros
product to the corresponding action with the Moyal product, using
\eqn{Tmoyvor}, to find which  Moyal theory corresponds to it.

The action \eqn{freeaction}is mapped into:
\bea
S'_0&=&\int \dd t \dd^2z\, T^{-1}({{\cal L}_0})\nn\\
&=& \int \dd t \dd^2z\,
\frac12\left(\left(\e^{-\frac\theta4\nabla^2}\del_\mu\varphi
\right)\starmoy\left(\e^{-\frac\theta4\nabla^2}\del_\mu\varphi\right)\right.\nonumber
\\&&\left.
-m^2\left(\e^{-\frac\theta4\nabla^2}\varphi\right)
\starmoy\left(\e^{-\frac\theta4\nabla^2}\varphi\right)\right)\nonumber\\
&=&\int \dd t \dd^2z\,
\frac12\left(\left(\e^{-\frac\theta4\nabla^2}\del_\mu\varphi\right)
\left(\e^{-\frac\theta4\nabla^2}\del_\mu\varphi\right)
-m^2\left(\e^{-\frac\theta4\nabla^2}\varphi\right)\left(
\e^{-\frac\theta4\nabla^2}\varphi\right)\right)\nonumber\\&=&\int
\dd t \dd^2z\,
\frac12\left(\del_\mu\varphi\e^{-\frac\theta2\nabla^2}\del_\mu\varphi
-m^2\varphi\e^{-\frac\theta2\nabla^2}\varphi\right)
\eea
which is not the free action with the Moyal product. In fact in
the latter case the noncommutative product could be eliminated
from the integral leaving just the free commutative action.
Therefore, the two actions being different,
 they could in principle give different equations of
motion.

Since we are dealing with a theory involving an infinite number of
derivatives we can ask whether we would need an infinity of
boundary conditions to solve the classical theory. This is not so,
as the higher derivatives appear as analytic functions of the
Laplacian, and in this case the boundary problem is the same as in
the standard case. Note that with our choice of $\theta^{\mu\nu}$
our equation is second order in the time derivatives, so that the
initial value problem requires knowledge of the field and its
derivative as initial condition. But also if we had deformed the
time derivatives, the initial data for the Cauchy problem would
have been the same if the higher derivative had been an analytic
function of the d'Alembertian. For more details and references see
the recent paper~\cite{BarnabyKamran}.

Let us derive the classical equations of motion starting from the
variation of the action. Since the standard techniques have been
developed for a theory with a finite number of derivatives, we
will proceed from first principles and start from the
infinitesimal variation of the field:
\begin{equation}
\varphi\to\varphi+\delta\varphi.
\end{equation}
It is not difficult to see that the corresponding infinitesimal
variation of the action under such a transformation is given by
\bea
\delta S_0&=&\int\mathrm{d}t \mathrm{d}^{2}z
\left((\partial_{0}\varphi)\starvor(\partial_{0}\delta\varphi)-
(\del_+\varphi)\starvor(\del_-\delta\varphi)-
(\del_+\delta\varphi)\starvor(\del_-\varphi)\nonumber\right.\\
&&\left.-m^{2}\varphi\starvor\delta\varphi\right),
\eea
where we have used the trace property of the integrals with
Wick-Voros products. By integrating by parts and using once again
the trace property we obtain, up to boundary terms:
\begin{equation}
\delta S_0=-\int\mathrm{d}t \mathrm{d}^{2}z(\delta\varphi)\starvor
\left(\partial_{0}\partial_{0}\varphi-2\del_+\del_-\varphi+m^{2}\varphi\right),
\end{equation}
namely
\begin{equation}
\delta S_0=-\int\mathrm{d}t \mathrm{d}^{2}z
\sum_{n=0}^{\infty}\frac{\theta^{n}}{n!}
\del_+^{n}(\delta\varphi)\del_-^{n}
\left(\partial_{0}\partial_{0}\varphi-2\del_+\del_-\varphi+m^{2}\varphi\right).
\end{equation}
By integrating once again by parts we obtain:
\begin{equation}
\delta S_0=-\int\mathrm{d}t \mathrm{d}^{2}z
\sum_{n=0}^{\infty}\frac{(-\theta)^{n}}{n!}(\delta\varphi)
\del_+^{n}\del_-^{n}
\left(\partial_{0}\partial_{0}\varphi-2\del_+\del_-\varphi+m^{2}\varphi\right),
\end{equation}
that is
\begin{equation}
\delta S_0=-\int\mathrm{d}t \mathrm{d}^{2}z
(\delta\varphi)\e^{-\theta\del_+\del_-}
\left(\partial_{0}\partial_{0}-2\del_+\del_-+m^{2}\right)\varphi.
\end{equation}
Since the variation of the action $\delta S$ must be vanishing for
any variation of the field $\delta\varphi$, we obtain that the
equation of motion is given by
\begin{equation}
\e^{-\theta\del_+\del_-}
\left(\partial_{0}\partial_{0}-2\del_+\del_-+m^{2}\right)\varphi=0.
\end{equation}
Equivalently, it can be written as:
\begin{equation}\label{EOM}
\e^{-\frac{\theta}{2}\nabla^{2}}\left(\Box+m^{2}\right)\varphi=0
\end{equation}
As we can see, the equation of motion \eqref{EOM} differs from the
classical Klein-Gordon equation
\begin{equation}\label{KGE}
\left(\Box+m^{2}\right)\varphi=0,
\end{equation}
only by the presence of the exponential of the Laplacian, an
invertible operator. It is immediate to see that all solutions of
the commutative theory are still solutions of the noncommutative
one. It is in principle possible that there can be  solutions of
the noncommutative equation \eqn{EOM} which are not solutions of
the commutative one. This is not the case, due to the
invertibility of the operator $\e^{-\frac{\theta}{2}\nabla^{2}}$ .

Notice that the on shell condition is not altered by the presence
of the deformation factor. In other words the dispersion relation
is the same in the two cases. In fact in Fourier transform
\begin{equation}
\varphi(x)=\int\frac{\mbox{d}^{3}k}{(2\pi)^3} e^{\ii k\cdot
x}\tilde{\varphi}(k)
\end{equation}
and then the equation \eqref{EOM} becomes:
\be
\e^{-\frac{\theta}{2}\nabla^{2}}\left(\Box+m^{2}\right)
\int\frac{\mbox{d}^{3}k}{(2\pi)^3} \e^{\ii k\cdot
x}\tilde{\varphi}(k)=\int\frac{\mbox{d}^{3}k}{(2\pi)^3}\,
\e^{\frac{\theta}{2}k^{2}}\left(-k^{2}+m^{2}\right)\e^{\ii k\cdot
x}\tilde{\varphi}(k)=0.
\ee
The relation
\begin{equation}
\e^{\frac{\theta}{2}k^{2}}\left(k^{2}-m^{2}\right)\tilde{\varphi}(k)=0,
\end{equation}
gives the same on shell relation of the classical case since the
exponential never vanishes.

Classically therefore the two theories have the same solutions of
the equations of the motion, despite the fact that the action, the
Lagrangian and the equations of motion are different.

\section{Green's Functions  \label{se:compare}}
\setcounter{equation}{0}

Let us consider a field theory described by the action:
\be
S=S_0+\frac{g}{4!}\int\dd t\dd^2 z\,
\varphi\star\varphi\star\varphi\star\varphi \label{action4}
\ee
where $\star$ is either $\starmoy$ or $\starvor$. In the following
we will use a generic $\star$ for all relations and formulas valid
for both products. We now calculate the Feynman rules for these
field theories.

Because of property~\eqn{Moyintegral} the free theory is unchanged
for the Moyal case. Therefore the Moyal propagator is the same as
in the undeformed case. In the Wick-Voros
case~\cite{ChaichianDemichevPresnajder} there are differences.

To this purpose let us rewrite the action $S_0$ in Eq.~\eqn{freac}
in the form
\be
S_0=\int \dd t \dd^2z\,\dd t' \dd^2z'\, \varphi(t,z)
K(t,t',z,z')\varphi(t',z')
\ee
with
\be
K(t,t',z,z')=\e^{-\frac{\theta}{2}\nabla^{2}}(-\del_\mu^2-m^2)
\delta(t-t') \delta^2(z-z').\label{kappa}
\ee

The quantum propagator $\Delta_{\starvor}(x,y)$  is the two-point
Green's function of the free theory, that is, the inverse of the
operator $K$,
\be
\Delta_{\starvor}(x_a,x_b)=\int\frac{\dd^3p}{(2\pi)^3} \e^{\ii
p\cdot(x_a-x_b)} \frac{\e^{-\frac\theta2|\vec
p|^2}}{p^2-m^2}.\label{propspacetime}
\ee
We can read off the propagator in momentum space, and compare it
with the one in the Moyal (and undeformed) case
\bea
{G^{(2)}_{0_M}}(p)&=&\frac{1}{p^2-m^2}\nonumber\\
{G^{(2)}_{0_V}}(p)&=&\frac{\e^{-\frac\theta2|\vec p|^2}}{p^2-m^2}
\label{propagators}
\eea
Since the poles in the propagator in momentum space are the same
as in the commutative theory, despite the change in the
propagator, the free field theory with the Wick-Voros product is
the same as in the commutative (and Moyal noncommutative) case.
This is the quantum counterpart of the previous result that the
solutions of the classical equations of the motion are the same.
Nevertheless the two propagators are not identical, and we will
have to take this into account in the following. Note however that
for infinite momentum there is an essential singularity, or a
zero, of the propagator, according to the sign of $\theta$. The
meaning of the essential singularity is not clear, but the oddity
is that the sign of $\theta$ has no physical meaning since it can
be changed by an exchange of the sign of one of the two
coordinates, in a theory which appears to be parity invariant. We
will see later that, with the proper twisting of the theory, also
this paradox is solved.

We now proceed to the calculation of the interaction vertex in the
Wick-Voros case, comparing it with the theory obtained with the
Moyal product. In this latter case the difference with respect to
the commutative case resides in the fact that the vertex acquires
a phase~\cite{Filk}. In order to see the corrections let us write
down the Moyal product as a convolution twist in momentum
space\footnote{Some of the formulas of this section are specific
to our 2+1 case, but the results are more general.}:
\be
(f\starmoy g)( x)=\int
\frac{\dd^3k}{(2\pi)^3}\,\frac{\dd^3k'}{(2\pi)^3}\tilde f( k)
\tilde g( k') \e^{\ii( k+ k')\cdot x}\e^{-\frac\ii2\theta\vec
k\wedge\vec k'} \label{MoyalprodFourier}
\ee
where $\tilde f$ and $\tilde g$ are the Fourier transforms of $f$
and $g$ and
\be
\vec k\wedge\vec k'=\varepsilon^{ij}k_i k'_j
\ee
We see that in momentum space the Moyal product is the standard
convolution of Fourier transforms, twisted by a phase. For the
Wick-Voros product, defining $k_\pm=(k_1\pm \ii k_-)/\sqrt{2}$  in
a way analogous to \eqn{defz} we have
\be
(f\starvor g)(z_+,z_-,t)=\int
\frac{\dd^3k}{(2\pi)^3}\,\frac{\dd^3k'}{(2\pi)^3}\tilde f( k)
\tilde g( k') \e^{\ii( k+k')\cdot x}\e^{-\theta k_- k'_+}
\ee
Explicitly the exponent of the twist in the convolution can be
expressed as
\be
k_- k'_+=\frac12\left(\vec k\cdot\vec k' +\ii\vec k\wedge\vec
k'\right)
\label{kprod}
\ee
%
with the same imaginary part  as in the Moyal case
\eqn{MoyalprodFourier} plus a real part.

For a $\varphi^4$ theory in ordinary space the four points vertex
in momentum space is the coupling constant multiplying the
$\delta$ of momentum conservation:
\be
V=-\ii \frac{g}{4!} (2\pi)^3\delta^3\left(\sum_{a=1}^4
{k_a}\right)
\ee
In the Moyal case we have that the vertex acquires a phase factor
due to the twist in the product~\eqn{MoyalprodFourier}:
\be
V_{\starmoy}=V\prod_{a<b}\e^{-\frac\ii2\theta^{ij}{k_a}_i{k_b}_j}
\ee
The presence of the phase in the vertex makes it non invariant for
a generic exchange of the momenta. This is a consequence of
noncommutativity and of the fact that the integral of Moyal
product of more than two functions is not invariant for an
exchange of the functions. Invariance for a cyclic rotation of the
factors still survives. This gives rise to a difference between
planar and nonplanar graphs, and ultimately to the well known
phenomenon of infrared-ultraviolet mixing~\cite{MvRS}.

In the Wick-Voros case the correction, due to \eqn{kprod}, is not
just a phase, but it contains a real exponent:
\be
V_{\starvor}=V\prod_{a<b}\e^{-\theta{{k_a}_-}{{k_b}_+}}=V\prod_{a<b}\e^{-\frac{\theta}{2}(
{ \vec {k_a}}\cdot{\vec{k_b}}+i {\vec{k_a}}\wedge{\vec{k_b}})}
\label{vertexVoros}
\ee
The exponent can have both signs, and in some case it could diverge
exponentially for large external momenta. The divergence is however
compensated by the fact that, to the four points function, there
must be added the contribution of the four propagators, each of
which comes with an exponentially convergent part. These convergent
parts compensate the possibly divergent contributions of the vertex
for positive $\theta$.

We can write the vertices with an unified notation as
\be
V_{\star}=V\prod_{a<b}\e^{\fase{k_a}{k_b}}
\ee
where
\be
\fase{k_a}{k_b}=\left\{\begin{array}{ll}
-\frac\ii2\theta^{ij}{k_a}_i{k_b}_j& \mbox{Moyal}\\
-\theta{{k_a}_-}{{k_b}_+} & \mbox{Wick-Voros}
\end{array}\right. \label{fase}
\ee

To calculate the four points Green's function in the Wick-Voros
case at the tree level we must attach to the vertex four
propagators~\eqn{propagators}, each carrying an exponential. The
four points Green's function therefore is
\be
G^{(4)}_{0_V}=-\ii {g} (2\pi)^3
\frac{\e^{-\theta\left(\sum_{a=1}^4
{k_a}_-{k_a}_++\sum_{a<b}{k_a}_-{k_b}_+\right)}}
{\prod_{a=1}^4{(k_a^2-m^2)}}\delta^{(3)}\left(\sum_{a=1}^4
k_a\right)
\ee
With some simple algebraic passages we can express the exponent as
\bea
&\sum_{a=1}^4{k_a}_-{k_a}_++\sum_{a<b}{k_a}_-{k_b}_+&\\
&=\frac14\left(|\vec k_1|^2 +|\vec k_2|^2+|\vec k_3|^2+|\vec
k_4|^2+2\ii\sum_{a<b}\vec k_a\wedge\vec k_b+|\vec k_1+\vec
k_2+\vec k_3+\vec k_4|^2\right)&\nonumber
\eea
The $\delta$ of conservation of momentum effectively kills the
last term, so that the four point function, at tree level, is
\be
G^{(4)}_{0_V}=-\ii {g} (2\pi)^3
\frac{\e^{-\frac\theta4\sum_{a=1}^4{|\vec
k_a|^2}}\prod_{a<b}\e^{-\frac\ii2\theta^{ij}{k_a}_i{k_b}_j}}
{\prod_{a=1}^4{(k_a^2-m^2)}}\,\delta^{(3)}\left(\sum_{a=1}^4
k_a\right) \label{G4tree}
\ee
Noticing that in the Moyal case, because of antisymmetry, it
results $\fase{p}{p}=0$, we can express in the unified notation
the Green's functions as:
\be
G^{(4)}_0= -\ii {g} (2\pi)^3 \frac{\e^{\sum_{a\leq
b}\fase{k_a}{k_b}}}
{\prod_{a=1}^4{(k_a^2-m^2)}}\delta\left(\sum_{a=1}^4 k_a\right)
\ee

\begin{figure}[htbp]
\epsfxsize=4.5 in
\bigskip
\centerline{\epsffile{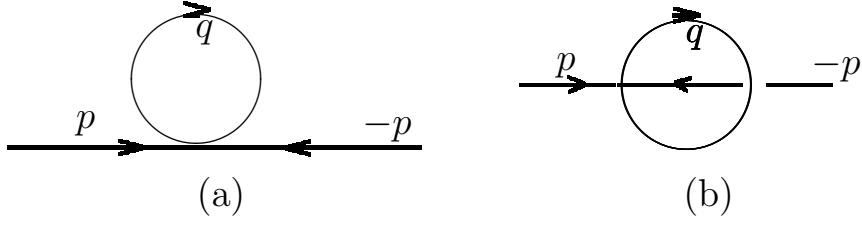}}
\caption{\sl The planar (a) and nonplanar (b) one loop correction
to the propagator} \label{planardiagram}
\end{figure}

The presence of a real exponent for the Wick-Voros case could
signify that the ultraviolet behaviour of the theory could be
different from the Moyal (and the commutative) case. Hence we
calculate the one loop correction to the propagator, and verify
the ultraviolet behaviour of the theory under renormalization. The
presence of the phase in the four point function in the complete
vertex~\eqn{G4tree} makes it non invariant for a generic
permutation of the external momenta, and this in turn implies that
the planar and nonplanar cases are to be treated differently, this
is what happens in the Moyal case as well. Consider first the
planar case in figure~\ref{planardiagram}~(a).
The amplitude is obtained using three
propagators~\eqn{propagators}, two with momentum $p$, one with
momentum $q$, and the vertex~\eqn{vertexVoros} with assignments
$k_1=-k_4=p$ and $k_2=-k_3=q$, and of course the integration in
$q$ and the proper symmetry factor:
\bea
G^{(2)}_{\mathrm P} &=& - \ii
\frac{g}{3}\int\frac{\dd^3q}{(2\pi)^3}\frac{\e^{-\theta(2p_-p_+
+q_- q_+)}\e^{-\theta(p_- q_+ - p_- q_+ - p_- p_+ -q_- q_+ -q_-
p_+ +q_- p_+)}}{(p^2-m^2)^2(q^2-m^2)}
\nonumber\\
&=&  - \ii \frac{g}{3}
\int\frac{\dd^3q}{(2\pi)^3}\frac{\e^{-\theta p_- p_+
}}{(p^2-m^2)^2(q^2-m^2)}
\eea
where the first exponential is due to the propagators, and the
second to the vertex. In this case all the contribution due to $q$
cancel, so that there is no change in the convergence of the
integral.

We now proceed to the discussion of  the nonplanar case, in
Fig.~\ref{planardiagram}~(b). The structure is the same as in the
planar case, but this time the assignments are instead
$k_1=-k_3=p$ and $k_2=-k_4=q$, and we have:
\bea
G^{(2)}_{\mathrm NP} &=& - \ii \frac{g}{6}
\int\frac{\dd^3q}{(2\pi)^3}\frac{\e^{-\theta(2p_- p_+ +q_-
q_+)}\e^{-\theta(p_- q_+ -p_- p_+ - p_- q_+ -q_- p_+ -q_- q_+ +
p_- q_+)}}{(p^2-m^2)^2(q^2-m^2)}
\nonumber\\
&=& - \ii
\frac{g}{6}\int\frac{\dd^3q}{(2\pi)^3}\frac{\e^{-\theta(p_- p_+
+\ii \vec p\wedge \vec q)}}{(p^2-m^2)^2(q^2-m^2)}
\eea
This time the $q$ contribution does not cancel completely, and
there remains the factor
\be
p_- q_+ - q_- p_+= \ii \vec p \wedge \vec q
\ee
so that the phase factor of the Moyal case is reproduced. We can
express in the unified notation:
\bea
G^{(2)}_{\mathrm P} &=& - \ii \frac{g}{3}
\int\frac{\dd^3q}{(2\pi)^3}\frac{\e^{\fase{p}{p}
}}{(p^2-m^2)^2(q^2-m^2)}\nonumber\\ G^{(2)}_{\mathrm NP} &=& - \ii
\frac{g}{6}
\int\frac{\dd^3q}{(2\pi)^3}\frac{\e^{\fase{p}{p}+\fase{p}{q}-
\fase{q}{p}}}{(p^2-m^2)^2(q^2-m^2)}
\eea

The ultraviolet divergence of the diagram is unchanged, with
respect to the commutative case, for the planar diagram. In the
nonplanar case there is the presence of the oscillating factor
$\ii \vec p \wedge \vec q$ in the exponential. This factor softens
the ultraviolet divergence, since it dampens the functions for
high $q$, but is responsible for infrared divergences. We can
conclude at this level that, while the Green's functions are
different, between the Moyal and Wick-Voros case, their
ultraviolet behaviour is the same as far as the momentum in the
loop is concerned\footnote{The convergence properties can however
be changed by going to a different noncommutative space, such as a
torus~\cite{ChaichianDemichevPresnajder}}. This indicates that the
noncommutative geometry, at the ultraviolet level, is basically
described by the uncertainty principle, consequence of the
commutator~\eqn{commutatorx}, which is unchanged between
Wick-Voros and Moyal cases. Nevertheless the two Green's functions
are not the same because of the $\fase{p}{p}$ term which vanishes
in the Moyal case, but not in the Wick-Voros one.

\begin{figure}[htb]
\epsfxsize=2.0 in
\bigskip
\centerline{\epsffile{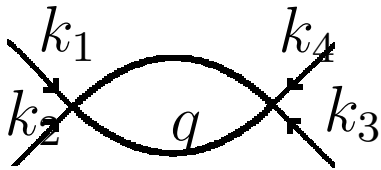}}
\caption{\sl The planar one loop four points diagram.}
\label{planarfish}
\end{figure}

We now proceed to the one-loop Green's functions corresponding to
the planar case of Fig.~\ref{planarfish}
In the NC case the Green's function correspondent to it can easily
be calculated by properly joining two vertices. It turns out that
we have for the two cases
\be
G^{(4)}_\mathrm{P}=\frac{(-\ii {g})^2}{8} (2\pi)^3\int\frac{\dd^3
q}{(2\pi)^3}\frac{\e^{\sum_{a\leq
b}\fase{k_a}{k_b}}\delta\left(\sum_{a=1}^4 k_a\right)}
{(q^2-m^2)((k_1+k_2-q)^2-m^2)\prod_{a=1}^4{(k_a^2-m^2)}}
\ee
The exponent in the numerator can be rewritten as in~\eqn{G4tree}
and we see that the internal momentum $q$ appears only in the
denominator, so that also in this case the planar diagram has the
same renormalization property of the undeformed theory. In the
Moyal case the real part exponent of the numerator is not present.
In the Wick-Voros case instead there is the same correction
encountered at tree level.

\begin{figure}[htb]
\epsfxsize=5.0 in
\bigskip
\centerline{\epsffile{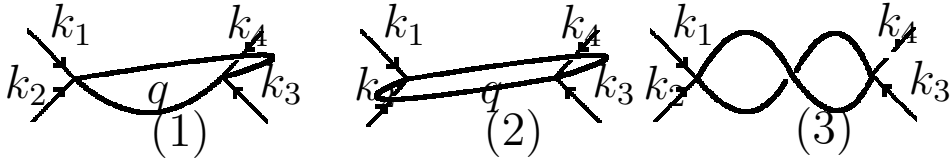}}
\caption{\sl The non planar one loop four points diagrams.}
\label{nonplanarfish}
\end{figure}

The three nonplanar cases are shown in figure~\ref{nonplanarfish}.
The calculation of their contribution is straightforward and
gives, in momentum space,
\bea
G^{(4)}_\mathrm{P}&=&\frac{(-\ii {g})^2}{8} \int\dd^3
q\frac{\e^{\sum_{a\leq b}\fase{k_a}{k_b}}\delta\left(\sum_{a=1}^4
k_a\right)}
{(q^2-m^2)((k_1+k_2-q)^2-m^2)\prod_{a=1}^4{(k_a^2-m^2)}}\nonumber\\\label{G4planar}\\
 G^{(4)}_\mathrm{{NP}_a}&=&(\frac{(-\ii {g})^2}{8} \int\dd^3 q
\frac{\e^{\sum_{a\leq
b}\fase{k_a}{k_b}+E_a}\delta\left(\sum_{a=1}^4 k_a\right)}
{(q^2-m^2)((k_1+k_2-q)^2-m^2)\prod_{a=1}^4{(k_a^2-m^2)}}\nonumber\\
\label{G4nonplanar}
\eea
with
\bea
E_1&=& \fase{q}{k_1}-\fase{k_1}{q} = \ii \vec q \wedge \vec k_1\nonumber\\
E_2&=& \fase{k_2}{q}-\fase{q}{k_2}+\fase{k_3}{q}-\fase{q}{k_3} =
\ii (\vec k_2 \wedge \vec q + \vec k_3 \wedge \vec q )\nonumber\\
E_3 &=&\fase{k_1}{q}-\fase{q}{k_1}+\fase{k_2}{q}-\fase{q}{k_2} =\ii
(\vec k_1 \wedge \vec q + \vec k_2 \wedge \vec q )\label{Ea}
\eea

Contrary to our expectations we find that the Green's functions
are different. The Green's functions are not however directly
measurable quantities, the S-matrix is. We will calculate it in
the twist-deformed framework in Sect.~\ref{se:smatrix}.

\section{The Wick-Voros and Moyal Products as Twisted
Noncommutative Geometries \label{se:twistncg}}
\setcounter{equation}{0}

The main physical motivation to study field theory equipped with a
$\star$ product is the belief that, at very short distances, the
geometry of spacetime is deformed, with the deformation dictated
by a small parameter, $\theta$ in our case. In the presence of
noncommutativity the concept of point is not well defined, and in
fact the proper mathematical formalism should use the theory of
$C^*$-algebras and the language of spectral triples (see for
example~\cite{connes, landi, ticos}). A star product deforms the
commutative algebra of functions on a space into a noncommutative
algebra. The proper formal definition of the mathematical objects
involved in the definition is beyond the scope of this article.
For us it suffices to know that the plane equipped Moyal product
can be made into a spectral triple~\cite{Selene, GG-BISV}.

As we have discussed in section~\ref{se:classical} the two products
come from a different quantization of the same Poisson structure,
which classifies $\star$ products up to
equivalences~\cite{Kontsevich}. They can also be seen as gauge
equivalent for the (infinite rank) group of gauge transformations
given by field redefinition of the kind~\eqn{equivalencemap}. Note
however that the action is not invariant under the action of this
gauge group.

With the introduction of a different, but equivalent, product one
can heuristically reason as follows. The presence of the
noncommutativity described by~\eqn{commutatorx} gives the
noncommutative structure of space, regardless of the realization
of the product one uses. In fact one could avoid the use of a
$\star$ product altogether, by considering the fields to be
infinite matrices  function of the operators $X$ defined
in~\eqn{defX} and solving, for example with a path integration,
this matrix model. We tested this conjecture for a bosonic quantum
field theory with a $\varphi^4$ interaction and found that the two
deformations of the action give different Green's functions.
Interestingly however the ultraviolet structure of the two
theories remains the same. We are nevertheless in front of a
puzzle.

The element that we need consider to solve this puzzle is
symmetries. The commutation relation~\eqn{commutatorx} breaks
Poincar\'e symmetry, and this is not a desirable feature for a
fundamental theory. The symmetry can be reinstated however at a
deformed level, considering the fact that both products can be
seen as coming from a Drinfeld twist~\cite{Drinfeld1,Drinfeld2}.
The noncommutative geometry described by either $\star$-product is
therefore a \emph{twisted} noncommutativity.

Given the Poincar\'e Lie algebra $\Xi$ and its universal
enveloping algebra~$U\Xi$, the twist $\mathcal{F}$ which we will
consider  is an element of $ U\Xi\otimes U\Xi.~$ For the Moyal and
Wick-Voros case it is respectively
\beqa
\mathcal{F}_M&=&\exp[-\ii\frac{\theta^{ij}}2 \del_i\otimes\del_j] \label{moytwist}\\
\mathcal{F}_V&=&\exp[-\theta\del_+\otimes\del_-] \label{vortwist}
\eeqa
where partial derivatives stand for translation generators and
have to be appropriately realized when acting on a given space.
Following~\cite{ADMW,ABDMSW,AschieriCorfu, mozart} we will
consider the following point of view: \emph{the noncommutative
geometry is a consequence of a twist of all products of the
theory}. Then every bilinear map~$\mu$ defined as
\be
\mu\,: X\otimes Y\rightarrow Z
\ee
(where $X,Y,Z$ are vector spaces)  is consistently deformed by
composing it with the appropriate realization of the twist
$\mathcal{F}$. In this way we obtain a deformed version
$\mu_\star$ of the initial bilinear map $\mu$:
\be
\mu_\star:=\mu\circ \mathcal{F}^{-1}~,\label{generalpres}
\ee
The $\star$-product on the space of functions is recovered setting
$X=Y=Z=\mathrm{Fun}(M)$. That is, if we indicate with $m_0$ the
usual pointwise product  between functions \footnote{At this level
we need not specify which kind of algebra of functions we are
considering. The algebra of formal series of the generators is
adequate, but more restricted algebras such as Schwarzian
functions can also be considered.}:
\bea
m_0: &&{\rm Fun}(M)\otimes {\rm Fun}(M) \longrightarrow {\rm Fun}((M)\nonumber\\
&& m_0(f\otimes g)=f\cdot g
\eea
the noncommutative product can be seen as the composition of $m_0$
with the twist:
\be
\mathcal{F}: {\rm Fun}(M)\otimes {\rm Fun}(M) \longrightarrow {\rm
Fun}(M)\otimes {\rm Fun}(M)
\ee
so that
\bea
f\starmoy g &=& m_0\circ \mathcal{F}^{-1}_{\starmoy} (f\otimes g)\nonumber\\
f\starvor g &=& m_0\circ \mathcal{F}^{-1}_{\starvor} (f\otimes g)
\eea

Associativity of the product is ensured by normalization and
cocycle conditions (see~\cite{AschieriCorfu,ADMW} for a short
introduction; see also the book~\cite{Majidbook}). We also
introduce the universal ${\mathcal R}$-matrix which represents the
permutation group in noncommutative space
\be
\mathcal R:= {\mathcal F}_{21} \mathcal{F}^{-1}
\ee
with
\be\mathcal{F}_{21}(a\otimes b ) = \tau\circ\mathcal{F}\circ\tau(a\otimes
b)
\ee
and  $\tau$  the usual exchange operator
\be
\tau(a\otimes b)=b\otimes a.
\ee
For the cases at hand with   the two twists given by
\eqn{moytwist}, \eqn{vortwist}:
 it results:
\be
\mathcal{R}_{\starvor}=\mathcal{R}_{\starmoy}=\mathcal{F}_{\starmoy}^{-2}
\ee
that is, the exchange operator, and therefore the statistics, are
the same in the two cases.

The presence of the twist deforms the structure of the universal
enveloping algebra of the Poincar\'e Lie-algebra, rendering it a
noncocommutative Hopf algebra. The analysis of~\cite{Helsinki,
Wess, aletWess}, made for the Moyal case, can be repeated in the
Wick-Voros case with similar conclusions. Therefore the
representations of the undeformed Poincar\'e algebra can still be
used. We will see later in the paper the important role that the
twisted Poincar\'e symmetry will play in the equivalence between
Moyal and Wick-Voros theories.

\section{Twist-deformed Products \label{se:twistproducts}}
\setcounter{equation}{0}

We have now the necessary ingredients to calculate a physical
process, like the S-matrix for the elastic scattering of two
particles. We recall than one of the crucial ingredients in the
importance of the S-matrix in physics is the issue of Lorentz and
Poincar\'e invariance. If we naively insert the Green's functions
of Sect.~\ref{se:compare} into the calculation of the S-matrix we
would find a dependence of it from the external momenta, something
like a momentum dependence of the coupling constant. What is more
relevant for our purposes, we find that the result would be
different for the Moyal and the Wick-Voros case, in contradiction
with the heuristic reasoning we made in the introduction. We would
also find a breaking of Poincar\'e invariance\footnote{We are
considering $\theta^{ij}$ to be constant. Another solution which
preserves Poincar\'e invariance is to have it a
tensor~\cite{DoplicherFredenhagenRoberts,BDFP} or to have it
transform together with the product~\cite{vega}. The residual
rotational invariance is an artifact of the two-dimensionality of
the model. In higher dimensions also this invariance is broken.}.

The reason for the breaking of Poincar\'e invariance is that the
commutator~\eqn{commutatorx} apparently breaks this invariance.
However the invariance can be reinstated if one  considers it to be
a \emph{quantum symmetry}, i.e.\ the Poincar\'e algebra is not a
cocommutative Hope algebra, but it has a nontrivial
coproduct~\cite{Helsinki,Wess}.

\begin{figure}[htb]\begin{center}
\epsfxsize=1.0 in
\bigskip
\centerline{\epsffile{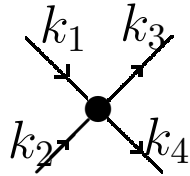}}
\caption{\sl The two particles elastic scattering} \label{Fig2to2}\end{center}
\end{figure}

Our purpose is to show, with an explicit calculation of scattering
amplitudes, that the naive procedure which leads to a difference
among the two cases can be corrected by a coherent twisting
procedure. We will see that, if the twisted symmetry is properly
implemented, the final, ``physical'' result, will be the same in
the Wick-Voros and Moyal cases, despite the presence of different
propagators and vertices. We will consider the elastic scattering
of two particles, as described in Fig.~\ref{Fig2to2}.

The first consequence of noncommutativity is the fact that, since
the vertex is non invariant for noncyclic exchange of the particles,
we have to twist-symmetrize the incoming and outgoing states using
the $\mathcal{R}$-matrix. Several aspects of this twist
symmetrization and the consequences for spin and statistics have
been discussed in~\cite{FioreSchupp1,FioreSchupp2,Balspinstat}.  In
the commutative case the order of the propagators into the vertex is
irrelevant, and therefore this discussion is redundant. Here there
are several twists at work and we have to be careful in considering
all of them.

Since we have to properly define multiparticle states as twisted
tensor products, and accordingly modify the definition of their
scalar product, for the remaining part of the section we will only
deal  with free fields solution of the Klein Gordon equation, and
free states. In the next section these will serve to define the
asymptotic states.

Let us consider the two-particle state:
\be
\ket{k_a,k_b}=\ket{k_a}\otimes\ket{k_b} \label{ketkakb}
\ee
Although for the comparison we are going to make later we will not
actually use the fact that the state has to be symmetrised, we will
discuss the symmetrisation of the states for completeness. Consider
the exchange operator
\be
\tau \ket{k_a}\otimes\ket{k_b}=\ket{k_b}\otimes\ket{k_a}
\ee
The symmetrized state, eigenvector of the exchange operator $\tau$
with eigenvalue~$+1$, is:
\be
\ket{k_a,k_b}_{simm}=\frac{\ket{k_a}\otimes\ket{k_b}+
\ket{k_b}\otimes\ket{k_a}}2 \label{ketkakbsimm}
\ee
inserting the two expressions~\eqn{ketkakb} or~\eqn{ketkakbsimm}
does not make a difference in the calculation (of the connected
diagrams) because of the invariance for exchange on the incoming
momenta. The symmetries for identical particles change for the
noncommutative case~\cite{FioreSchupp1,FioreSchupp2, Balspinstat},
we have to take into account the fact that the tensor product is
twisted, and moreover that the exchange is twisted as well.
Therefore we define
\be
\ket{k_a,k_b}_\star=\tilde{\mathcal{F}}^{-1}\ket{k_a,k_b}
\ee
where by $\tilde{\mathcal{F}}$ we indicate the twist that this time
acts in momentum space:
\bea
\tilde{\mathcal{F}}_{\starmoy}^{-1} \ket{k_a}\otimes\ket{k_b} &=&
\e^{-\frac\ii2\theta^{ij} k_{a_i}\otimes
k_{b_j}}\ket{k_a}\otimes\ket{k_b}
\nonumber\\
\tilde{\mathcal{F}}_{\starvor}^{-1}
\ket{k_a}\otimes\ket{k_b}&=&\e^{\theta k_{a_-}\otimes k_{b_+}}
\ket{k_a}\otimes\ket{k_b} \label{twist2ket}
\eea
This is not the only change we have to make to the
state~\eqn{ketkakbsimm}: the state has to be eigenvalue of the
twist-exchange, given by the $\mathcal{R}$-matrix acting in momentum
space. The properly symmetrized state is therefore
\bea
\ket{k_a,k_b}_{simm_\star} &=&
\frac12\left(\tilde{\mathcal{F}}^{-1}\ket{k_a}\otimes\ket{k_b} +
\tilde{\mathcal{F}}^{-1}\tilde{\mathcal{R}}^{-1}\ket{k_a}\otimes\ket{k_b}\right)\nonumber\\
&=&\frac12\left( \tilde{\mathcal{F}}^{-1}\ket{k_a}\otimes\ket{k_b}
+\tilde{\mathcal{F}}^{-1}\tilde{\mathcal{F}}\tilde{\mathcal{F}}^{-1}_{21}\ket{k_a}\otimes
\ket{k_b}\right)\label{twist2sym}
\eea
We define as usual the momentum eigenstates as created by the
creation operators $a_k, a_k^\dag$:
\be
\ket{k}=a^\dagger_k\ket{0} \label{ketk}
\ee
where $a_k, a_k^\dag$ are obtained in terms of the free field
\be
\varphi(x)=\int \frac{d^2 k}{\sqrt{(2\pi)^2 2\omega_k}} (a_k
\e^{-\ii k\cdot x}+a^\dag_k \e^{\ii k\cdot x})
\ee
  to be
\bea
a_k&=& \frac{\ii}{\sqrt{ {(2\pi)^2} 2 \omega_k}} \int d^2 x \e^{\ii
k \cdot x}
\stackrel{\leftrightarrow}{\del_0} \varphi_{in}(x) \nonumber\\
a^\dagger_k&=& -\frac{\ii}{\sqrt{{(2\pi)^2} 2 \omega_k}} \int d^2 x
\e^{-\ii k \cdot x} \stackrel{\leftrightarrow}{\del_0}
\varphi_{in}(x)\label{expansiona}
\eea
 The operators $a_k, a_k^\dag$ may be regarded, for
fixed $k$, as functionals of the fields, therefore their $\star$
product may be obtained as in~\cite{mozart}
\bea a(k)\starmoy a(k')
&=&\e^{-\frac{\ii}{2}\theta^{ij}k_ik'_j}\,a(k)a(k')\nonumber\\
a(k)\starmoy a^\dag(k')
&=&\e^{\frac{\ii}{2}\theta^{ij}k_ik'_j}\,a(k)a^\dag(k')
\nn\\
a^\dag(k)\starmoy a(k')
&=&\e^{-\frac{\ii}{2}\theta^{ij}k_ik'_j}\,a^\dag(k)a(k')
\label{starak1}
\eea
\bea a(k)\starvor a(k')
&=&\e^{-\theta k_- k'_+}\,a(k)a^\dag(k')\nonumber\\
a(k)\starvor a^\dag(k') &=&\e^{\theta k_-k'_+}\,a(k)a^\dag(k')\nn\\
a^\dag(k)\starvor a(k') &=&\e^{-\theta k_-k'_+}\,a^\dag(k)a(k').
\label{starak2}
\eea
Therefore we may reexpress Eqs.~\eqn{twist2ket} and \eqn{twist2sym}
as
\be \ket{k_a,k_b}_{\star}=a^\dagger_{k_a}\star
a^\dagger_{k_b}\ket{0} \label{tensorkakb}
\ee
and
\be
\ket{k_a,k_b}_{simm_\star}=\frac{a^\dagger_{k_a}\star
a^\dagger_{k_b}+a^\dagger_{k_b}\star a^\dagger_{k_a}}2\ket{0}
\label{tensorkakbsym}
\ee
  The next step is the twist of the inner
product. We consider it as a map from two copies of the Fock space
of states into complex numbers. In the commutative case, for the
momentum one particle states we have:
\be
\hs{\cdot}{\cdot}:\, \ket{k}\otimes\ket{k'}\longrightarrow
\hs{k}{k'}=\bra{0}a_ka^\dagger_{k'}\ket{0}=\delta({k-k'})
\ee
We twist this product in the usual way composing it with the twist
operator:
\bea
\hsstar{\cdot}{\cdot}:\, \ket{k}\otimes\ket{k'}\longrightarrow
  \hs{\cdot}{\cdot}\circ \mathcal{F}^{-1}:\,
\ket{k}\otimes\ket{k'}&=&\tilde{\mathcal{F}}^{-1}(k,k')\hs{k}{k'}\nonumber\\&=&
\bra{0}a_k\star a^\dagger_{k'}\ket{0} \label{twistinner1part}
\eea
with $\tilde{\mathcal{F}}^{-1}(k,k')$ given by the exponential
factor in Eqs.~\eqn{twist2ket} for the Moyal and Wick-Voros case
respectively.

 We are not yet finished twisting! Let us consider the inner
product in the commutative case:
\be
\hs{k_1,k_2}{k_3,k_4}=\delta(k_1-k_3)\delta(k_2-k_4)
\ee
In the noncommutative case we have to twist  the two-particle state
according to~\eqn{twist2ket}, and then we have to twist the inner
product according to the two-particle generalization
of~\eqn{twistinner1part}. In order to realise such a generalization
we must consider the action of the twist on two-particle states.
This is done, in canonical form, via the coproduct of the Hope
algebra. Given a representation of an element of the Hope algebra on
a space, the representation of the element on the product of states
is given (in the undeformed case) by
\be
\Delta_0(u)(f\otimes g)=(1\otimes u + u\otimes 1)(f\otimes g)
\label{coprod0}
\ee
The coproduct is responsible for example of the Leibnitz rule. For
the twisted Hope algebra the  coproduct is deformed according to the
fact that it is the $\mathcal{R}$-matrix which realizes the
permutations:
\be
\Delta_\star(u)(f\otimes g)=(1\otimes u + \mathcal{R}^{-1}(u\otimes
1))(f\otimes g) \label{coprodstar}
\ee
However  the twists we are considering are built out of
translations, whose coproduct is undeformed:
\be
\Delta_0(\del_i)=\Delta_{\starmoy}(\del_i)=\Delta_{\starvor}(\del_i)=1\otimes
\del_i + \del_i\otimes 1
\ee
Since we are acting on two-particle states we need to define also
\be
\Delta_0(\del_i\otimes\del_j)=\Delta_{\star}(\del_i\otimes\del_j)=1\otimes
1\otimes \del_i \otimes\del_j+ \del_i\otimes\del_j\otimes 1 \otimes
1
\ee
Therefore the twisted inner product among two-particle states
\be
\hsstar{k_1 k_2}{k_3 k_4}=
\hs{\cdot}{\cdot}\circ\Delta_\star({\mathcal{F}}^{-1})(\, \ket{k_1
k_2}\otimes\ket{k_3 k_4})
\ee
may be easily computed to be
\bea
\hsstarmoy{k_1,k_2}{k_3,k_4}&=&\e^{\frac\ii 2
\theta^{ij}(k_{1_i}+k_{2_i})(k_{3_j}+k_{4_j})}
\hs{k_1,k_2}{k_3,k_4}\nonumber\\
\hsstarvor{k_1,k_2}{k_3,k_4}&=&\e^{
\theta({k_1}_-+{k_2}_-)({k_3}_++{k_4}_-)} \hs{k_1,k_2}{k_3,k_4}
\label{twistinner2}
\eea
We can now calculate the \emph{twisted} inner product of
\emph{twisted} states. Combining \eqn{twistinner2} with
\eqn{twist2ket} we obtain  the simple expression:
\bea
{\phantom{\bigg\rangle}}_{\starmoy}\!\!\!\hsstarmoy{k_1,k_2}
{k_3,k_4}_{\!\!\!\starmoy}&=&\e^{\frac\ii 2 \theta^{ij}\sum_{a<b}
k_{a_i}k_{b_j}}
\hs{k_1,k_2}{k_3,k_4}\nonumber\\
{\phantom{\bigg\rangle}}_{\starvor}\!\!\!\hsstarvor{k_1,k_2}
{k_3,k_4}_{\!\!\!\starvor}&=&\e^{ \theta\sum_{a<b}{k_a}_-{k_b}_+}
\hs{k_1,k_2}{k_3,k_4}\label{twistinnerprod}
\eea
that is
\be
{\phantom{\bigg\rangle}}_{\star}\!\!\!\hsstar{k_1,k_2}
{k_3,k_4}_{\!\!\!\star}= \e^{-\sum_{a< b}\fase{k_a}{k_b}} \hs{k_1,k_2}{k_3,k_4}
\ee
with $\fase{k_a}{k_b}$ defined in \eqn{fase}.

Recalling the results~\eqn{starak1} and~\eqn{starak2}, we can cast
the previous expression in the form:
\be
{\phantom{\bigg\rangle}}_{\star}\!\!\!\hsstar{k_1,k_2}{k_3,k_4}_{\!\!\!\star}
=\bra{0}a_{k_1}\star a_{k_2} \star a_{k_3}^\dagger \star
a_{k_4}^\dagger \ket{0}
\ee
This is in some sense also a consistency check. We could have
started with the commutative expression
$\hs{k_1,k_2}{k_3,k_4}=\bra{0}a_{k_1} a_{k_2} a_{k_3}^\dagger
a_{k_4}^\dagger \ket{0}$ and twisted the product among the creation
and annihilation operators, obtaining the above result. We decided
to follow a longer procedure to highlight the appearance of the
various twists.

\section{The twisted S-matrix \label{se:smatrix}}
\setcounter{equation}{0}

Let $\ket{f}, \ket{i}$ denote a collection of free asymptotic states
at $t=\pm\infty$ respectively. We also assume that we can define in
some way the one particle incoming and outgoing states. This is a
nontrivial assumption\footnote{We thank Harald Grosse for bringing
this fact to our attention.}, in a theory in which localization is
impossible the concept of localization may not be well defined.
Nevertheless is it reasonable to expect that also in this theory,
for small $\theta$ for large distances and times it will be possible
to talk on incoming and outgoing states, expandable in the
eigenvalues of momentum $\ket{k}$.

As in standard books in quantum field theory we define the S matrix
as the matrix which describes  the scattering of the initial $\ket
i$ states into the final $\ket f$ states
\be
S_{fi}={\phantom{\bigg\rangle}}_{ in \star}\left\langle
{f}\stackrel{\star}{\big|}{ i} \right\rangle_{\star out
}={\phantom{\bigg\rangle}}_{ out \star}\left\langle
{f}\stackrel{\star}{\big|}S\stackrel{\star}{\big|} { i}
\right\rangle_{\star out}={\phantom{\bigg\rangle}}_{ in
\star}\left\langle
{f}\stackrel{\star}{\big|}S\stackrel{\star}{\big|} { i}
\right\rangle_{\star in}
\ee
 where the twisted inner
product of twisted states \eqn{twistinnerprod} is understood. The
one-particle asymptotic state is defined as in \eqn{ketk} to be
\be
\ket{k}_{in}=N_\star(k) a^\dagger_k\ket{0}_{in}=-N_\star (k)\frac{\ii}{\sqrt{{(2\pi)^2} 2
\omega_k}} \int d^2 x \e^{-\ii k \cdot x}
\stackrel{\leftrightarrow}{\del_0}
\varphi_{in}(x)\ket{0}_{in}\label{kek}
\ee
with $N_\star(k)$ a normalization factor to be determined for the
Moyal and Wick-Voros cases separately. Analogously for the out
states. Moreover we assume, as in the commutative case, that the
matrix elements of the interacting field $\varphi(x)$ approaches
those of the free asymptotic field as time goes to $\mp \infty$
\be
\lim_{x^0\rightarrow\pm\infty} \langle f|\varphi(x)|i\rangle=Z^{1/2}
\langle f|\varphi_{\stackrel{out}{in}}(x)|i\rangle \label{freint}
\ee
with $Z$ a renormalization factor.
 To be
definite let us consider an elastic process of two particles in two
particles. According to the previous section we have then
 \be
{S_{fi}}_\star (k_1,..,k_4)={\phantom{\bigg\rangle}}_{ in \star}\hstar{k_1 k_2}{k_3
k_4}_{in \star }= \e^{\sum_{a< b}\fase{k_a}{k_b}}
{\phantom{\rangle}}_{ in }\hs {k_1 k_2}{  k_3 k_4}_{ out}
\ee
which can be expressed in terms of Green's functions, following
the same procedure as in the commutative case (see for
example~\cite{Kaku}).  On repeatedly using \eqn{kek} and
\eqn{freint} we arrive at
\beqa
S_{fi}&=&{\phantom{\bigg\rangle}}_{ in \star}\hstar{k_1 k_2}{k_3
k_4}_{out \star }= {\rm disconnected}\;\; {\rm graphs}\nn\\
&+&\bar N_\star(k_1)
\bar N_\star(k_2)N_\star(k_3)N_\star(k_4){(iZ^{-1/2})^2}\e^{-\sum_{a< b}
\fase{k_a}{k_b}} \label{ourproc}\\
&\times& \int \frac{\Pi_a\; d^2 x^a} {\sqrt{(2\pi)^2
2\omega_{k_a}}}\e^{-\ii k_a x^a} (\del_\mu^2 +m^2)_a
G(x_1,x_2,x_3,x_4)\nn
\eeqa
where $G(x_1,x_2,x_3,x_4)$ is the four-point Green's function.

In order to fix the normalization of the asymptotic states let us
compute the scattering amplitude for one particle going into one
particle, at zeroth order. Up to the undeformed normalization
factors $N(p_a)$, this has to give a delta function
\bea
\bar N(k) N(p) \delta^2(k-p)&=&N_\star^*(k)
N_\star(p){\phantom{\bigg\rangle}}_{ in \star}\hstar{k}{p}_{out
\star }\nonumber\\ &=& N_\star^*(k) N_\star(p) \e^{\fase{-k} {p}}
{\phantom{\rangle}}_{ in }\hs {k}{ p}_{ out}\nonumber\\
&=&N_\star^*(k) N_\star(p) \e^{\fase{-k} {p}} \delta^2(k-p)
\eea
from which
\beqa
N_{\starmoy}(p)&=&N(p)\nn\\
N_{\starvor}(p)&=&\e^{-\frac{\theta}{4}|\vec p|^2} N(p).
\eeqa

Let us now compute the scattering amplitude for the process above
(two-particles  in two particles) at one loop. We have two kinds of
contribution to~\eqn{ourproc}, one coming from  the planar
terms~\eqn{G4planar}, which in spatial coordinates read
\be
G_\mathrm{P}(x_1,x_2,x_3,x_4)= \int \Pi_a \frac{d^2
k_a}{\sqrt{(2\pi)^2 2\omega_{k_a}}} \e^{i k_a x^a}
G^{(4)}_\mathrm{P} (k_1,k_2,k_3,k_4)
\ee
the other coming from non planar terms \eqn{G4nonplanar}
\be
G_\mathrm{NP}(x_1,x_2,x_3,x_4)= \int \Pi_a \frac{d^2
k_a}{\sqrt{(2\pi)^2 2\omega_{k_a}}} \e^{i k_a x^a}
G^{(4)}_\mathrm{NP} (k_1,k_2,k_3,k_4)
\ee

Let us do the computation for the planar case first. Substituting in
\eqn{ourproc} we find the same result in Moyal and Wick-Voros case;
moreover they coincide with the undeformed result:
\beqa
&&{S_{fi}}_{\star P}(k_1,..,k_4)= \frac{(-\ii g)^2}{8}(2\pi)^3 \bar N(k_1) \bar N(k_2) N(k_3)
N(k_4) \Pi_a\e^{\frac{\theta}{4}|\vec k_a|^2} \nn\\
&& \e^{-\sum_{a<b}\fase{k_a}{k_b}} \int\Pi_a \frac{d^2
x^a}{\sqrt{(2\pi)^2 2\omega_{k_a}}} \e^{-\ii k_a x^a}\int \Pi_a
\frac{d^2 p_a}{\sqrt{(2\pi)^2 2\omega_{p_a}}} \e^{\ii p_a
x^a}(-p_a^2+m^2) \nn\\&& \int\frac{\dd^3 q}{(2\pi)^3} \frac{\e^{\sum_{a\leq
b}\fase{p_a}{p_b}}\delta\left(\sum_{a=1}^4 p_a\right)}
{(q^2-m^2)((p_1+k_2-q)^2-m^2)\prod_{a=1}^4{(p_a^2-m^2)}}
\eeqa
The integration over the $x^a$ variables yields factors of $(2\pi)^2
\delta^{(2)}(k_a-p_a)$; therefore the propagators of the external
legs cancel as in the standard case, as well as the factor
\be
\Pi_a\e^{\frac{\theta}{4}|\vec
k_a|^2}\e^{-\sum_{a<b}\fase{k_a}{k_b}} \times\e^{\sum_{a\leq
b}\fase{p_a}{p_b}}\delta^{(2)}(k_a-p_a) \longrightarrow
1\label{canc}
\ee
 so that we are left with the usual {commutative} expression
\be
{S_{fi}}_{\star P}(k_1,..,k_4)=
 S_{fi}(k_1,..,k_4)
 \ee
In the NP case instead we find
\beqa
&&{S_{fi}}_{\star NP}(k_1,..,k_4)= \frac{(-\ii g)^2}{8}(2\pi)^3 \bar N(k_1) \bar N(k_2) N(k_3) N(k_4)
\Pi_a\e^{\frac{\theta}{4}|\vec k_a|^2} \nn\\
&& \e^{-\sum_{a<b}\fase{k_a}{k_b}} \int\Pi_a \frac{d^2
x^a}{\sqrt{(2\pi)^2 2\omega_{k_a}}} \e^{-\ii k_a x^a}\int \Pi_a
\frac{d^2
p_a}{\sqrt{(2\pi)^2 2\omega_{p_a}}} \e^{\ii p_a x^a}(-p_a^2+m^2) \nn\\
&& \int\frac{\dd^3 q}{(2\pi)^3} \frac{\e^{\sum_{a\leq
b}\fase{p_a}{p_b}+E_a}\delta\left(\sum_{a=1}^4 p_a\right)}
{(q^2-m^2)((p_1+k_2-q)^2-m^2)\prod_{a=1}^4{(p_a^2-m^2)}}
\eeqa
After integrating over $x^a$ the propagators of the external legs
cancel and the simplification~\eqn{canc} continues to hold, but we
are left with the exponential of $E_a$ which does not simplify.
Its explicit expression is given in \eqn{Ea}, as we can see it is
an imaginary phase, and it has \emph{ the same expression in the
Moyal and Wick-Voros case}. It depends on the $q$ variable,
therefore it gets integrated and modifies the IR and UV behaviour
of the loop: this is the correction responsible for the UV/IR
mixing~\cite{MvRS}. Therefore we can conclude that
\be
{S_{fi}}_{\starmoy NP}(k_1,..,k_4)={S_{fi}}_{\starvor NP}(k_1,..,k_4)\ne{S_{fi}}(k_1,..,k_4)
\ee

\section{Conclusions}
In Giuseppe Tomasi di Lampedusa's novel \emph{Il Gattopardo}
(translated \emph{The Leopard})~\cite{TomasidiLampedusa} the
Prince of Salina says: ``Change everything so that nothing
changes". This sums up the situation that we faced in our analysis
of the field theory in the presence of the Wick-Voros and Moyal
products. We started with different actions, coming from different
Lagrangians densities already at the level of the free theory. The
free propagator for the Wick-Voros is different from the Moyal
case, but the classical theory has no new solutions, and at the
quantum level the poles of the propagators are the same. Then we
found a different vertex for the quartic theory, which led to a
different four points function. But the differences are reabsorbed
in the S-matrix, \emph{provided one recognizes the properly
normalized asymptotic states}. It is not anymore enough to think
of a flux of particles to be identified by ordinary plane waves
described by the usual exponential wave with the customary
dispersion relation. In a field theory with a different propagator
such as the one considered here the asymptotic states change. The
noncommutative cases are however different from the commutative
one (something has to change!), but they describe the same
``physics'' among themselves.

The two noncommutative products (Moyal and Wick-Voros) are
different realizations of the same algebra, and as such describe
the same noncommutative geometry, and it would have been curious
to find different physical consequences. In fact one could have
studied the noncommutative geometry exclusively at the operatorial
level, without the need for a deformed product. But at the end of
the day, to confront with a physical theory, one has to map the
states into physically observable states, that an experimenter (at
least an ideal one) can prepare. The correspondence between states
and real objects is not immediate in noncommutative geometry, and
has to be handled with extreme care. This is the morale of this
tale. In noncommutative geometry, the different structure of space
time forces to change the correspondence between mathematical
objects and physical observables. This should lead to the
formulation of a coherent theory of observables and measurements
on noncommutative spaces.

\subsubsection*{Acknowledgments}
We thank A.P. Balachandran, Seckin Kurkcuoglu, Renato Musto, Denjoe
O'Connor and Pietro Santorelli for discussions and correspondence.

\end{document}